\documentclass[aps,prl,twocolumn,floatfix,superscriptaddress]{revtex4-2}

\usepackage{bm,bbm,amsmath, amssymb,amsbsy, amsthm,graphicx,subfigure,color,txfonts,multirow,enumitem} 
\usepackage[framemethod=tikz]{mdframed}
\usepackage[export]{adjustbox}
\usepackage[colorlinks]{hyperref}
\hypersetup{
	bookmarksnumbered,
	pdfstartview={FitH},
	citecolor={blue},
	linkcolor={blue},
	urlcolor={blue},
	pdfpagemode={UseOutlines}
}
\usepackage{cleveref}

\newcommand{\ket}[1]{|#1\rangle}
\newcommand{\bra}[1]{\langle#1|}

\newcommand{\tr}[1]{\mathrm{tr}\left\{#1\right\}}
\newcommand{\ptr}[2]{\mathrm{tr_{#1}}\left\{#2\right\}}
\newcommand*\sq{\mathbin{\vcenter{\hbox{\rule{.3ex}{.3ex}}}}}

\begin{document}
 
 \title{Redundantly amplified information suppresses  quantum correlations in many-body systems}
 


 \author{Davide Girolami}
\email{davegirolami@gmail.com}
 \affiliation{$\hbox{DISAT, Politecnico di Torino, Corso Duca degli Abruzzi 24, Torino 10129, Italy}$
 }
\author{Akram Touil} 
\affiliation{$\hbox{Department of Physics, University of Maryland, Baltimore County, Baltimore, MD 21250, USA}$
 }
 \affiliation{$\hbox{Center for Nonlinear Studies, Los Alamos National Laboratory, Los Alamos, New Mexico 87545}$}
 \author{Bin Yan}
  \affiliation{$\hbox{Center for Nonlinear Studies, Los Alamos National Laboratory, Los Alamos, New Mexico 87545}$}
\affiliation{$\hbox{Theoretical Division, Los Alamos National Laboratory, Los Alamos, New Mexico 87545}$
 }
 \author{Sebastian Deffner}
\affiliation{$\hbox{Department of Physics, University of Maryland, Baltimore County, Baltimore, MD 21250, USA}$
 }
 \affiliation{$\hbox{Instituto de Fisica ‘Gleb Wataghin’, Universidade Estadual de Campinas, 13083-859, Campinas, Sao Paulo, Brazil}$}
 \author{Wojciech H. Zurek}
\affiliation{$\hbox{Theoretical Division, Los Alamos National Laboratory, Los Alamos, New Mexico 87545}$
 }
\begin{abstract} 
We establish  bounds on quantum correlations in many-body systems. They reveal what sort of information about a quantum system can be simultaneously recorded in different parts of its environment. Specifically, independent agents who monitor environment fragments can eavesdrop only on amplified and redundantly disseminated -- hence, effectively classical -- information about the decoherence-resistant pointer observable. We also show that the emergence of classical objectivity is signaled by a distinctive scaling of the conditional mutual information, bypassing hard numerical optimizations. Our results validate the core idea of Quantum Darwinism: objective classical reality does not need to be postulated and is not accidental, but rather a compelling emergent feature of quantum theory  that otherwise  
-- in absence of decoherence and amplification -- leads to ``quantum weirdness''. In particular, a lack of consensus between agents that access environment fragments is bounded by the information deficit, a measure of the incompleteness of the information about the system. 
\end{abstract}

\date{\today}

\maketitle

\paragraph{Introduction.} Is classical reality, reflected in the consensus between independent agents about the properties of physical systems \cite{deco}, a consequence of quantum laws?  Quantum weirdness makes it difficult to reconcile human perception with our most successful scientific theory. 
In particular, quantum systems display stronger correlations than those admitted by classical physics \cite{EPR,epr2,entanglement}. They enable the advantages of quantum information processing \cite{telep,super,capa,naturedisc}. 
Despite their importance in  quantum science, our understanding of genuinely quantum correlations is limited: their identification and quantification  in large scale quantum systems---the focus of  quantum-classical transition---is an often intractable problem \cite{tech,corr1,corr2,discorev,sanpera,newzurek,sen}. 

Here, we prove universal, quantitative bounds on quantum correlations  in many-body systems: they are bounded by the shared classical information among their parts. As an important consequence, objectivity of measurement results  arises only when quantum correlations between an information source and a network of recipients are selectively suppressed. That is, consensus responsible for objective classical reality is an emergent attribute of Quantum Mechanics.  

 First, we consider a quantum universe consisting of a system ${\cal S}$ and an environment ${\cal E}$. We prove an upper bound on   quantum discord, which quantifies genuinely quantum correlations \cite{discordzurek}. The simultaneous creation of quantum discord between ${\cal S}$ and different environment fragments ${\cal F}$ and ${\cal E}/{\cal F}$ is restricted. 
 The upper limit is determined  by how much classical information about ${\cal S}$ is concurrently available to observers monitoring the two distinct fragments. 

Then, we extend our study to the multipartite case.  Quantum correlations are generally {\it not} monogamous, and almost ubiquitous in Hilbert space \cite{koashi,mono,braga,monogamy,acin}. Nevertheless, we prove an upper bound on the {\it average} bipartite quantum discord, and, remarkably, also on the entanglement of formation that can exist between ${\cal S}$ and any of $N$  subsystems $\varepsilon_i$  of the environment. Simultaneous classical correlations between ${\cal S}$ and each $\varepsilon_i$ imply that quantum discord (almost) vanishes throughout the universe. Hence, quantum information about ${\cal S}$ is inaccessible to independent  observers that monitor different $\varepsilon_i$. 

This result supports Quantum Darwinism, pinpointing the origin of classical reality within quantum theory \cite{darwin1}. Its core insight is that independent observers (such as humans) find out about ${\cal S}$ by eavesdropping on $\varepsilon_i$s -- e.g., scattered or emitted photons in our everyday  ${\cal E}$  \cite{darwin2,darwin4,darwin5,darwin6,exp1,exp2,pan}.  Only information that has been replicated throughout the environment \cite{Z07,Z13}, resulting in multiple  records, is widely accessible---only pointer states that survive decoherence 
 intact and can be shared   by many  observers become subject of consensus, acquiring a classically objective nature~\cite{Zurek81,deco}.

The newfound bounds on quantum correlations confirm that 
agreement among independent observers suppresses quantumness. 
Only large  
fragments (i.e., ${\cal F} \ge {\cal E}/{\cal F}$)
retain  quantum information about ${\cal S}$. Moreover, we show that when disjoint environment fragments establish sufficient correlations with ${\cal S}$, they store predominantly information about a \emph{unique} observable. We compute bounds on such classical correlations between environment fragments, obtaining an information-theoretic characterization of objective classical reality. These bounds generalize previous findings ~\cite{ollivier,ollivier2,brandao,ranard,adesso,ollivier3,riedelbranch,fu}, highlighting that redundancy of information available to independent observers implies uniqueness of objective reality. 

Finally, we introduce an analytical witness of objectivity.  Testing Quantum Darwinism in complex systems is hard, because quantifying correlations requires daunting numerical optimizations \cite{discome,corr1}.    We overcome this limitation and show that redundancy of classical correlations (in its strongest form) is signaled by a characteristic scaling of the conditional mutual information,  an analytical function of quantum states  \cite{cover}.
  
\begin{figure}
\includegraphics[width=.46\textwidth,height=4.5cm,fbox=0.5pt 5pt]{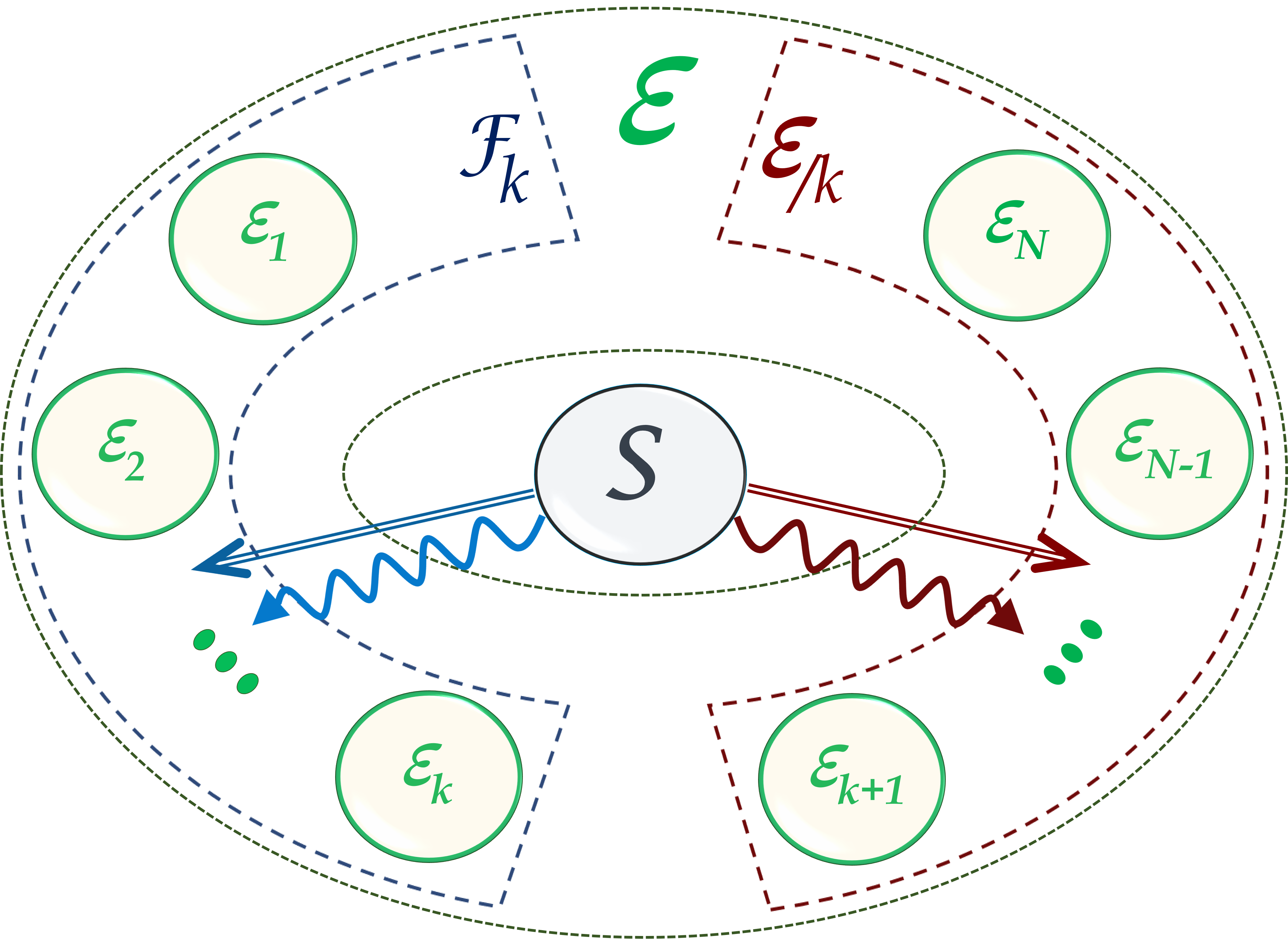}
\caption{We demonstrate quantitative bounds on  quantum    correlations between a system ${\cal S}$ and fragments  ${\cal F}_k,{\cal E}_{/k}$ of an $N$-partite environment ${\cal E}$. Equation~(\ref{eq4}) is an upper limit to  quantum discord   (wavy lines) in terms of the consensus about  classical  information (double lines) that is broadcast from ${\cal S}$ to ${\cal F}_k$ and    ${\cal E}_{/k}.$}
\label{fig1}
\end{figure}

\paragraph{Trade-off relations for quantum correlations.}  We consider a   quantum system ${\cal S}$ of dimension $d_{{\cal S}}$ and an $N$-partite environment  ${\cal E}:=\cup_{i=1}^N\varepsilon_i$ of dimension $d_{{\cal E}}=\Pi_{i=1}^N d_{\varepsilon_i}$. We call ${\cal F}_{k}:=\cup_{\# i=k}\varepsilon_i$ a fragment of  $k< N$ elements, and  ${\cal E}_{/k}:={\cal E}/{\cal F}_k$ its complement.   The  information shared by ${\cal S}$ and  ${\cal F}_{k}$ is quantified by the mutual information $I({\cal S}:{\cal F}_{k}):=H({\cal S})+H({\cal F}_{k})-H({\cal S}\,{\cal F}_{k}),$ where $H({\cal X}):=-\tr{\rho_{\cal X}\log_2 \rho_{\cal X}}\leq \log_2 d_{{\cal X}}$ is the von Neumann entropy of the state $\rho_{{\cal X}}$ of ${\cal X}$. The mutual information consists of classical and quantum components \cite{discordzurek,vedral}. The classical part is the (maximal) mutual information that is left after a local measurement  $\mathbf{M}_k:=\left\{\mathbf{M}_\alpha, \sum_\alpha \mathbf{M}_\alpha^\dagger\mathbf{M}_\alpha= \mathbb{I}_{d_{{\cal F}_k}}\right\}$ on ${\cal F}_{k}$.  Given the post-measurement state 
 \begin{equation}
 \rho_{{\cal SF}_{k,\mathbf{M}_k}}=\sum_\alpha \left(\mathbb{I}_{d_{{\cal S}}}\otimes \mathbf{M}_{\alpha}\right)\,\rho_{{\cal SF}_k} \left(\mathbb{I}_{d_{{\cal S}}}\otimes \mathbf{M}_\alpha^{\dagger}\right),
 \end{equation}
 classical correlations are quantified as the maximal information about ${\cal S}$ an observer can extract  by measurements on ${\cal F}_k$:  $J\left({\cal S}:\check{\cal F}_{k}\right):=\max_{\mathbf{M}_k} I\left({\cal S}:{\cal F}_{k,\mathbf{M}_k}\right)$ \cite{vedral,holevo}. This quantity is upper bounded by $H({\cal S})$. \emph{Quantum discord}, the most general kind of quantum correlation, is then defined as the difference between pre-measurement and post-measurement mutual information, 
 \begin{equation}
 D\left({\cal S}:\check{\cal F}_k\right):=I({\cal S}:{\cal F}_{k})-J\left({\cal S}:\check{\cal F}_{k}\right).
 \end{equation}
Note that classical and quantum correlations are generally not invariant under subsystem swapping: $J\left({\cal S}:\check{\cal F}_k\right)\neq J\left(\check{\cal S}:{\cal F}_k\right)$, and $D\left({\cal S}:\check{\cal F}_k\right)\neq D\left(\check{\cal S}:{\cal F}_k\right)$.

 Quantum discord  
 $D\left({\cal S}:\check{\cal F}_k\right)$ is the minimum {\it quantum information} about ${\cal S}$  that   ${\cal F}_k$ loses  when a local measurement $\mathbf{M}_{k}$ is performed \cite{streltsovzurek,wiseman,EPR}.
 Quantum discord can exist even in non-entangled states \cite{discordzurek,acin}, as it can be created by local operations and classical communication (LOCCs) \cite{locc}. Specifically, $D\left({\cal S}:\check{\cal F}_k\right)=0$ if and only if there exists a measurement $\tilde{\mathbf{M}}_k$ such that $\rho_{{\cal SF}_{k}}=\rho_{{\cal SF}_{k,\mathbf{\tilde{M}}_k}}$. 
 Quantum discord signals the presence of quantum coherence \cite{convert}. It can be converted into entanglement \cite{StreltsovBruss,adessopiani}, and it is a resource for quantum metrology \cite{lqu}. For pure states, it is  equal to the entanglement entropy, $D\left({\cal S}:\check{\cal F}_k\right)=D\left(\check{\cal S}:{\cal F}_k\right)=H({\cal S})$, while in general its maximal value is $H({\cal F}_k)$ \cite{tech}.   
 
 \begin{figure}
\includegraphics[width=.46\textwidth,height=4.5cm,,fbox=0.5pt 5pt]{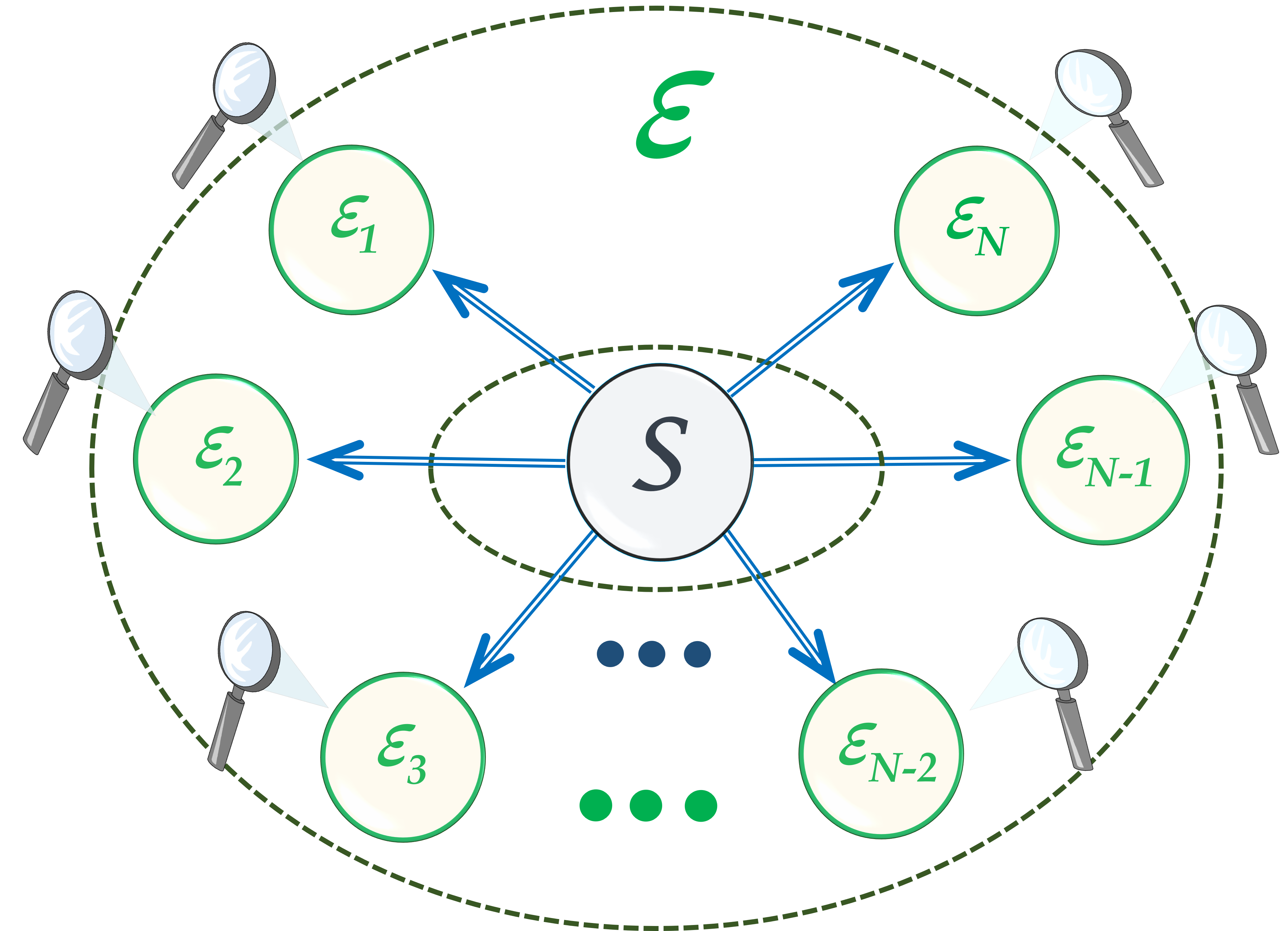}
\caption{Quantum Darwinism recognizes that information about a system  ${\cal S}$ is obtained from disjoint environment fragments consisting of distinct subsystems $\varepsilon_i$s of ${\cal E}$ by independent observers. Unconstrained proliferation of classical correlations means that only the information about a pointer observable is accessible.   Equation (\ref{main}) implies that, whenever classical objectivity manifests, bipartite quantum correlations are suppressed.}
\label{fig3}
\end{figure}

 \begin{figure*}
\subfigure{\fbox{\includegraphics[width=.32\textwidth,height=4cm]{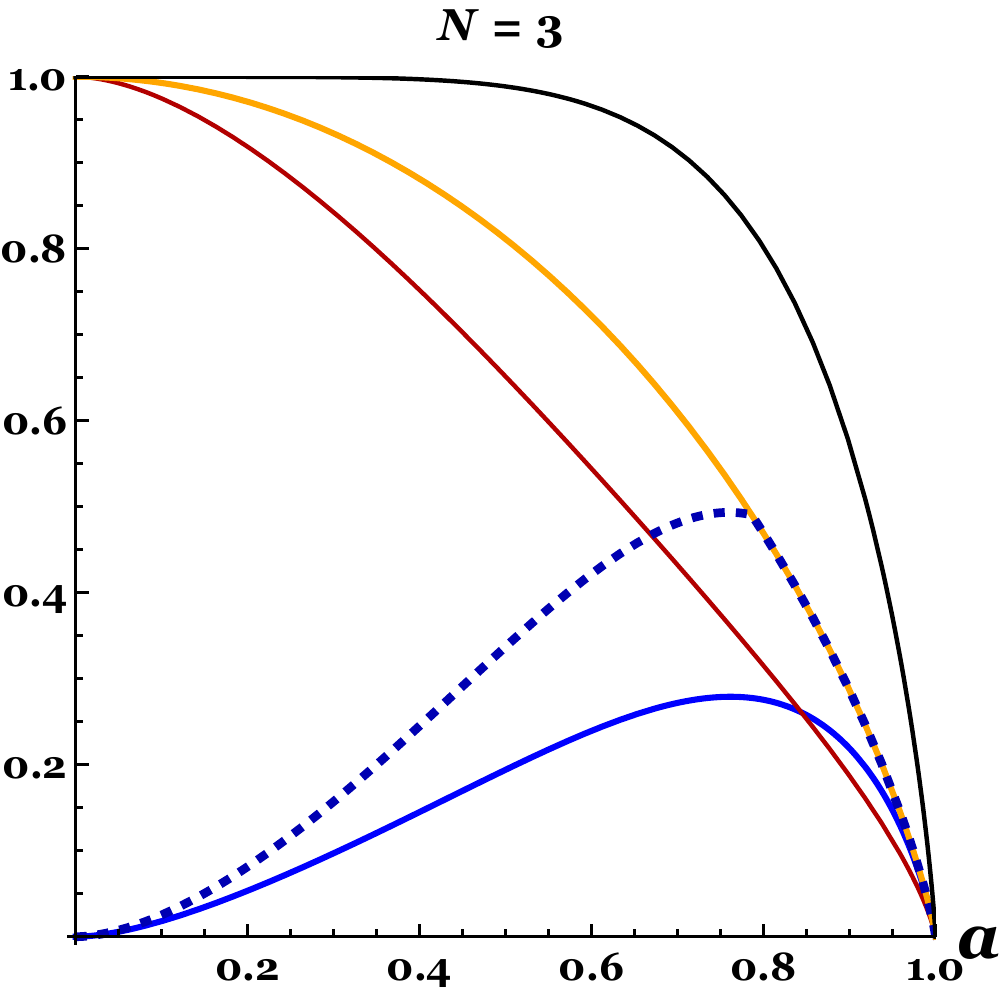}\hspace{5pt}
\includegraphics[width=.32\textwidth,height=4cm]{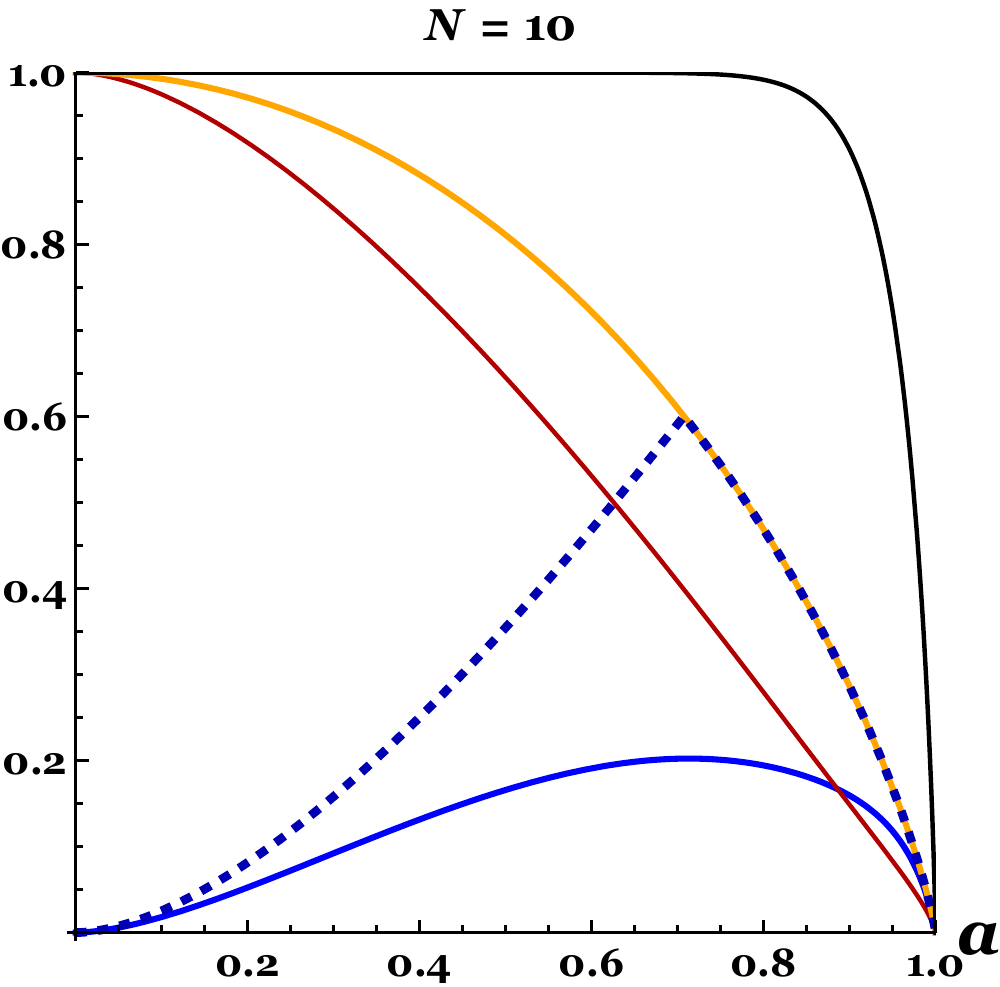}\hspace{5pt}
\includegraphics[width=.32\textwidth,height=4cm]{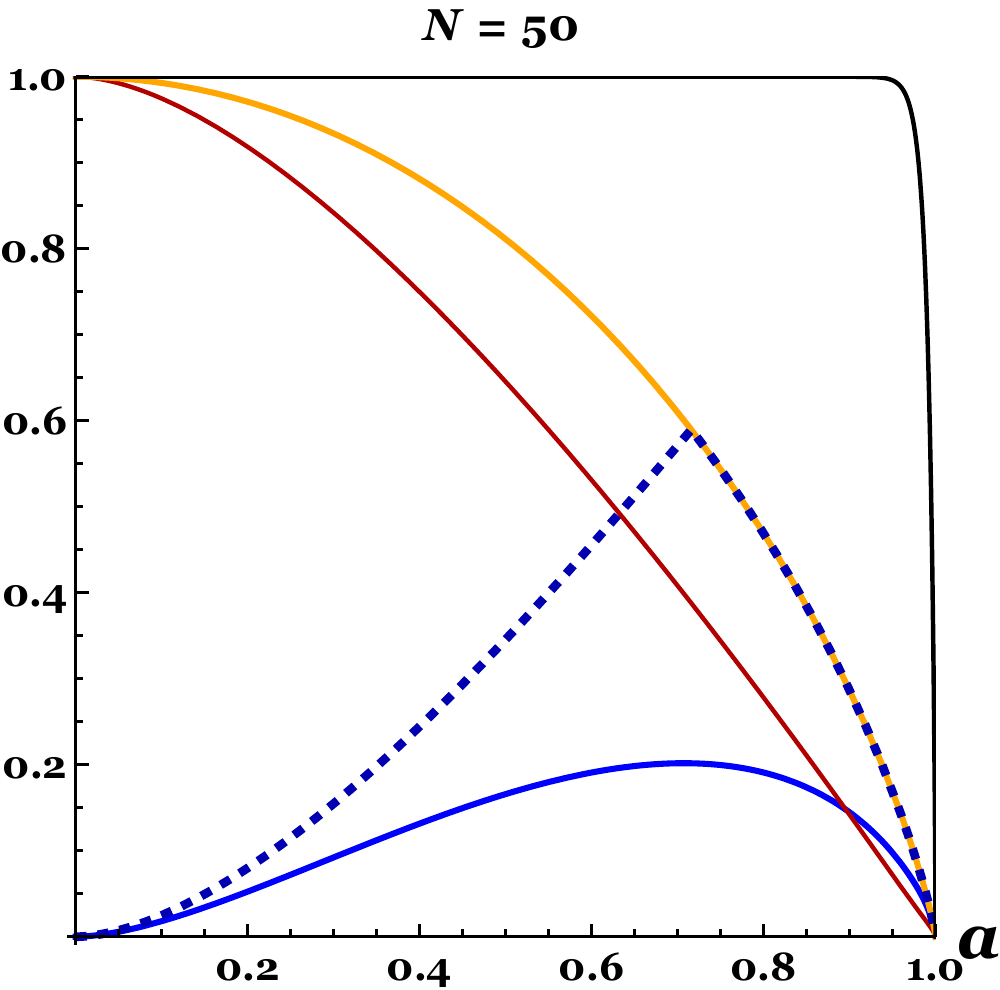}}}
\vspace{-10pt}
\caption{
By employing known methods  \cite{ranktwo,akram}, for different values of $N$, we compute  the  following quantities in the  state ${\bf U}_{{\cal SE}}(a)\ket{+}_{\cal S}\ket{0}_{\cal E}^{\otimes N}$  (\cite{epaps}):   quantum discord, $\bar D\left({\cal S}:\check{\varepsilon}_{i}\right)$ (blue line \textcolor{rgb:red,0;green,0;blue,0.5}{{\large ---}}); known upper bound $H(\varepsilon_i),$ (orange line \textcolor{orange}{{\large ---}});  minimum among the upper bound from Eq.~(\ref{main}) and $H(\varepsilon_i)$  (dashed blue line \textcolor{rgb:red,0;green,0;blue,0.5}{{\large$\sq\sq\sq$}});  classical correlations $\bar J\left({\cal S}:\check{\varepsilon}_{i}\right)$ (red line \textcolor{rgb:red,1;green,0.3;blue,0.3}{{\large ---}});  $H({\cal S})$ (black line {\large ---}).  For $N=2$, the bound in Eq.~(\ref{main}) is even saturated, $\bar D\left({\cal S}:\check{\varepsilon}_{i}\right)=\delta\,H({\cal S})$. Overall, it is much more informative than the entropic limit for $a\rightarrow 0$, while classical correlations attain $H({\cal S})$.}
\label{fig2}
\end{figure*}

In the following, we derive contraints on quantum correlations between ${\cal S}$ and any fragment ${\cal F}$. 
First, we evaluate upper bounds to $D\left({\cal S}:\check{\cal F}_k\right)$. That is, how much quantum information about ${\cal S}$ is accessible to an observer who knows the state of ${\cal F}_k$ (Fig.~\ref{fig1}). Koashi and Winter discovered a trade-off between the  entanglement of formation $E_f({\cal S}:{\cal F}_k)$ in ${\cal SF}_k$  \cite{notent},  and classical correlations in ${\cal SE}_{/k}$ \cite{entform,koashi},
\begin{equation}
\label{koashi}
\begin{split}
E_f({\cal S}:{\cal F}_k)\leq\, H({\cal S})-J\left({\cal S}:\check{\cal E}_{/k}\right).
\end{split}
\end{equation}
Without loss of generality, we assume now that  ${\cal SE}$  is in a pure state $\ket{\psi}_{{\cal SE}}$.  Then, the inequality in Eq.~(\ref{koashi}) is saturated.  This surprising relation between classical and quantum features  does not hold if we replace entanglement with quantum discord \cite{koashi,fanchini}. 
 
There is an exact bound on quantum discord  between ${\cal S}$ and environment fragments.  
We quantify the (lack of) agreement between the classical information about ${\cal S}$ that is accessible via ${\cal F}_{k}$ and ${\cal E}_{/k}$, i.e., {\it classical objectivity} \cite{ollivier2,korbiczreview}, by introducing the information deficit
\begin{eqnarray}\label{delta}
\delta:= \frac{
J\left({\cal S}:\check{{\cal E}}\right)-\min\left\{J\left({\cal S}:\check{{\cal F}}_k\right),J\left({\cal S}:\check{{\cal E}}_{/k}\right)\right\}}{H({\cal S})}\, \in\,[0,1].
\end{eqnarray}
The information deficit disappears if and only if  classical information about ${\cal S}$ is simultaneously stored into ${\cal F}_k$ and ${\cal E}_{/k}$, and it is maximal if and only if there is maximal discrepancy  \cite{epaps}. The information deficit $\delta$ was employed in previous Quantum Darwinism literature as a free parameter with no physical interpretation. Here, it has the crucial role of objectivity measure. The definition in Eq.~(\ref{delta}) is key in the proof that classical objectivity restricts proliferation of quantum correlations. 

{\it {\bf Result 1}: For any state of the universe $\ket{\psi}_{{\cal SE}}$},
\begin{equation}
\label{eq4}
D\left({\cal S}:\check{\cal F}_k\right) +D\left({\cal S}:\check{\cal E}_{/k}\right) \leq  
2\,\delta\,H({\cal S}).
\end{equation}
{\it Proof -- } Since $J\left({\cal S}:\check{{\cal E}}\right)=H({\cal S})$ for pure states  $\ket{\psi}_{{\cal SE}}$, one has
\begin{align*}
 I({\cal S}:{\cal F}_k)+ I({\cal S}:{\cal E}_{/k})&= 2\, H({\cal S})\Rightarrow\\
 D\left({\cal S}:\check{{\cal F}}_k\right)+ D\left({\cal S}:\check{{\cal E}}_{/k}\right)&= 2\, H({\cal S})- J\left({\cal S}:\check{{\cal F}}_k\right)-J\left({\cal S}:\check{{\cal E}}_{/k}\right)\\
   &\leq 2\, H({\cal S}) +2\,\delta\, H({\cal S})-2\, J\left({\cal S}:\check{{\cal E}}\right)\\
   &\leq 2\,\delta\, H({\cal S}).
 \end{align*}
Hence, consensus between two observers  accessing ${\cal F}_k,{\cal E}_{/k}$, respectively,  about classical information on ${\cal S}$ prevents proliferation of quantum correlations. Note that for $\delta\rightarrow0$, neither fragment can share quantum discord with ${\cal S}$.

We extend this result, by proving a bound on the concurrent sharing of quantum information about ${\cal S}$ with $N$ environment constituents $\varepsilon_i$, i.e., to $N>2$ observers (Fig.~\ref{fig3}).     
As a special case of Eq.~(\ref{delta}), we quantify   the (lack of) consensus of two observers accessing $\varepsilon_i$ and ${\cal E}_{/i}$ by 
\begin{align}
\delta_i:=\ \frac{
J\left({\cal S}:\check{{\cal E}}\right)-\min\left\{J\left({\cal S}:\check{\varepsilon}_i\right),J\left({\cal S}:\check{{\cal E}}_{/i}\right)\right\}}{H({\cal S})}\, \in\,[0,1].
\end{align}
It is also useful to define the average information deficit   $\delta:=\sum_{i=1}^{N} \delta_i/N$. For $N=2$, it is the quantity in Eq.~(\ref{delta}) \cite{notealt}. 
 Then, there exists a universal bound on quantum discord in many-body systems \cite{epaps}:
 
 {\bf Result 2:} {\it For any state of the universe $\ket{\psi}_{{\cal SE}}$,} 
  \begin{align}
  \label{main}
 \bar{D}\left({\cal S}:\check\varepsilon_{i}\right):=& \frac1N\sum_{i=1}^N  D\left({\cal S}:\check\varepsilon_{i}\right),\nonumber\\
  \bar{D}\left({\cal S}:\check\varepsilon_{i}\right)\leq&\,
\delta\,H({\cal S}).
 \end{align}
The tightness of this bound only depends on  the   $\delta$, i.e.,  (lack of) objectivity.  Also, since $E_f({\cal S}:\varepsilon_i)=H({\cal S})-J\left({\cal S}:\check{{\cal E}}_{/i}\right)$, and $\delta_i=1-
\min\left\{J\left({\cal S}:\check{\varepsilon}_i\right),J\left({\cal S}:\check{{\cal E}}_{/i}\right)\right\}/{H({\cal S})},$ the entanglement of formation is upper bounded: $E_f({\cal S}:\varepsilon_i)\leq \delta_i\,H({\cal S})$. Averaging over all subsystems $\varepsilon_i$, we get

{\bf Remark:} {\it For any  $\ket{\psi}_{{\cal SE}},$}
\begin{align}\label{entbound} 
\bar{E}_f\left({\cal S}:\varepsilon_{i}\right):=\frac1N\sum_{i=1}^N  E_f\left({\cal S}:\varepsilon_{i}\right)\leq\,
\delta\,H({\cal S}).
\end{align}
  That is,  quantum correlations are tightly constrained whenever multiple observers reach agreement concerning classical information about ${\cal S}$. 
 Note that the inequality  $\bar{J}\left({\cal S}:\check\varepsilon_{i}\right):= \frac1N\sum_{i=1}^N  J\left({\cal S}:\check\varepsilon_{i}\right) \leq H({\cal S})$ can be saturated: independent observers can achieve  arbitrarily small $\delta$, making  quantum correlations  (almost) vanish. \\
 We support this statement with an example. (See ~\cite{epaps} for details.)
  A qubit ${\cal S}$ and an $N$-qubit  environment ${\cal E}$ are in the initial state $\ket{+}_{{\cal S}} \ket{0}^{\otimes N}_{{\cal E}}$, with $\ket{+}\equiv \sqrt{1/2}(\ket{0}+\ket{1})$. We quantify classical and quantum correlations that are created by a unitary ${\bf U}_{{\cal SE}}(a)\equiv\Pi_{i=1}^N {\bf U}_{{\cal S}\varepsilon_i}(a)$, where each ${\bf U}_{{\cal S}\varepsilon_i}(a)$ implements the controlled gate $\mathbb{I}_2\oplus  \left( \begin{smallmatrix} a&\sqrt{1-a^2}\\ \sqrt{1-a^2}&-a \end{smallmatrix} \right), a\in[0,1],$ on ${\cal S}\varepsilon_i$.  Their average values, in this case, are the values calculated for any ${\cal S}\varepsilon_i$ bipartition. This dynamical ``c-maybe'' model \cite{akram} is significant: it can represent the correlation pattern of a system ${\cal S}$ interacting with a photonic environment. The universe is therefore in a {\it singly-branching} state~\cite{photon2,darwin2}. The plots in Fig.~(\ref{fig2})  highlight how the newfound bound to quantum discord  is much tighter than the entropic limit $H(\varepsilon_i)$ in the most interesting regime, when system and environment are highly correlated ($a\rightarrow 0$; the universe is in a (generalized) GHZ state). 
  
  While limits to quantum information sharing are manifest in GHZ states, the generality of Eqs. (\ref{main},\ref{entbound})  is surprising. Quantum discord and the entanglement of formation  are generally non-monogamous:   $D\left({\cal S}: \check{{\cal E}}\right)\ngeq \sum_iD\left({\cal S}: \check{\varepsilon}_{i}\right)$ \cite{koashi,fanchini,wootters,monogamy,mono,giorgi,braga}. Also, there are infinitely many kinds of entanglement structures, i.e., classes of states that cannot be transformed into each other by (stochastic) LOCCs \cite{cirac}. Our bounds therefore capture a  universal feature of many-body quantum systems which cannot be inferred from the structure of the GHZ class, nor by monogamy relations.

We stress that \emph{redundancy} of amplified -- hence, classical -- information is sufficient to suppress quantum correlations.
For pure states $\ket{\psi}_{{\cal SE}}$, 
 \begin{equation}
 \label{main2}
 \begin{split}
  \bar J\left({\cal S}:\check\varepsilon_{i}\right)\geq\, (1- \delta)\,H({\cal S})\Rightarrow
  \bar{D}\left({\cal S}:\check\varepsilon_{i}\right)\leq\, \delta\,H({\cal S}).
  \end{split}
 \end{equation}
Moreover, 
 if at least $(1- \delta)\,H({\cal S})$ bits of classical correlations are shared between ${\cal S}$ and  $R_\delta$ subsystems $\varepsilon_i$, then
  \begin{equation}\label{redundancy}
  \bar{D}\left({\cal S}:\check{{\cal F}}_k\right)\leq \left(1-R_\delta\,(1-\delta)/N\right)\,H({\cal S}),\,k\leq N/2,
  \end{equation}
  where the average is computed over all ${\cal F}_k$s. 
 When classical information is redundantly broadcast ($R_\delta \approx N,\, \delta\approx 0$), only large fragments ($k>N/2$)  display any traces of quantum correlations with ${\cal S}$.  

Recent works discovered bounds on entanglement sharing for generic ${\cal SE}$ dynamics  \cite{brandao,adesso,ranard}. Specifically, a state $\rho_{{\cal S}\varepsilon_{i}}=\ptr{{{\cal E}_{/i}}}{{\bf U}_{{\cal SE}}\ket{\psi}_{{\cal SE}}}$  is always close to a separable state, displaying zero discord in the ideal limit $N\rightarrow \infty$.  Here, we have obtained exact, physically meaningful bounds on quantum correlations assuming a realistic, finite environment. In particular, when observers accessing different ${\varepsilon}_i$s agree with each other (small $\delta$), quantum information is inaccessible, while classical information can spread into the environment.  
This is how consensus that defines objective classical reality  emerges from a quantum substrate \cite{ollivier}, as we discuss in the following.

\paragraph{Significance of the bounds within Quantum Darwinism.} 

 A physical state is  \emph{classically objective} if independent  observers agree about its properties. \emph{Quantum Darwinism} describes the origin of classical objectivity within quantum theory \cite{darwin1,ollivier}.  
 Different observers access information about a  system ${\cal S}$ by eavesdropping on different parts of the  environment (Fig.~\ref{fig3}). Because of decoherence, only information about ``pointer observables'' $\left\{\hat{\mathbf{M}}:=\ket{\hat\alpha}\bra{\hat\alpha}\right\}$ is communicated through the environment \cite{note}, e.g., by photons that interact with a central system and then carry information about it. 
 Such scattered light  \cite{photon2,photon} is then intercepted by rod cells or artificial photoreceptors.  Crucially, only classical information survives  decoherence and becomes  
 available to observers  \cite{darwin1,ollivier}. The statement is formalized by a characteristic scaling of classical correlations:  
\begin{equation}\label{otherdarwin}
J\left(\check{\cal S}:{\cal F}_{k}\right)\geq (1-\delta)\,H({\cal S}),  \,\forall\, {\cal F}_k,\,  k\geq k_{\delta}\,,
\end{equation}
in which $k_{\delta}\ll N$ is determined by the information deficit $\delta$ \cite{ollivier,darwin1,darwin4,darwin5,darwin6}.  That is, any fragment $\mathcal{F}_k$ carries the same large amount of classical information about ${\cal S}$. However, we have recently established that Quantum Darwinism can be better expressed by the scaling of 
 classical correlations with respect to measurements on $\mathcal{F}_k$ \cite{akram},
\begin{equation}
\label{eq7}
J\left({\cal S}:\check{\cal F}_k\right)\geq (1-\delta)\,H({\cal S}),\,\forall\, {\cal F}_k,\, k\geq k_{\delta}.
\end{equation}
  We stress that  $J\left({\cal S}:\check{\cal F}_k\right)$ is the maximal information  about ${\cal S}$ one extracts by measuring on  ${\cal F}_k$ \cite{noteclass}.

Our result Eq.~\eqref{main} corroborates   Quantum Darwinism's central tenet.  Recognizing the information deficit $\delta$ as a measure of (lack of) classical objectivity elucidates how  redundancy of classical information suppresses  quantum correlations. In particular, for pure states, Quantum Darwinism (Eq.~(\ref{eq7}))  implies
\begin{align}\label{discord}
 D\left({\cal S}:\check{\cal F}_{k}\right)\leq2\,\delta\,H({\cal S}),\,\forall\,{\cal F}_k,\, k\in[k_\delta,N-k_\delta],
 \end{align}
  certifying that quantum information is not concurrently accessible to multiple independent observers.  

 Further, we prove that Eq.~(\ref{eq7}), and therefore our bound on quantum discord, signify  uniqueness of the pointer observable  \cite{epaps}:
 
{\it {\bf Result 3:} For  any disjoint fragments ${\cal F}_{k},{\cal F}_l,$ and any state  ${\bf U}_{{\cal SE}_{/k+l}}{\bf V}_{{\cal SF}_l}{\bf W}_{{\cal SF}_k}\ket{\psi}_{{\cal S}}\ket{\phi}_{{\cal F}_k}\ket{\varphi}_{{\cal F}_l}\ket{\chi}_{{\cal E}_{/k+l}}$, $\,k,l\geq k_\delta$, if Eq.~(\ref{eq7}) holds, then}
\begin{align}
\label{eq10}
\left(1-2\,\delta\right)H({\cal S})&\leq  I\left({\cal F}_{k,\hat{\mathbf{M}}_k}:{\cal F}_{l,\hat{\mathbf{M}}_l}\right),\\
 I\left({\cal F}_{k,\hat{\mathbf{M}}_k}:{\cal F}_{l,\hat{\mathbf{M}}_l}\right)&\leq \begin{cases}
 \left(1+\,\delta\right)\,H({\cal S}),\,\,
    \text{if}\,\Delta_{{\bf U}_{{\cal SE}_{/k+l}}{\bf V}_{{\cal SF}_l}}H({\cal S})\geq 0,\\
    \log_2 d_{{\cal S}}+\delta\,H({\cal S}), \, \text{otherwise},
\end{cases}\nonumber
\end{align}
where $\Delta_{ {\bf U}_{{\cal SE}_{/k+l}}{\bf V}_{{\cal SF}_l}}H({\cal S})$ is the entropy variation due to ${\bf U}_{{\cal SE}_{/k+l}}{\bf V}_{{\cal SF}_l}$. The lower limit holds, in fact, for any pure state $\ket{\psi}_{{\cal SE}}$, while the restriction on the dynamics is necessary to establish the upper bound, as it ensures that classical correlations between fragments are strictly information about ${\cal S}$. These  stringent bounds show that the maximally informative observables in disjoint fragments  are  highly correlated, $\hat{\mathbf{M}}_{k}\approx \hat{\mathbf{M}}_{l}\approx\hat{\mathbf{M}}$. When ${\cal S}$ and small fragments ${\cal F}_{k}, k\geq k_\delta,$ already share maximal classical correlations, the maximally informative measurement for any observer is inevitably  the projection on the pointer basis $\{\hat \alpha\}$.  While  similar statements were proven for the ideal case of $\delta=0$  \cite{ollivier,ollivier2,riedelbranch,fu,ollivier3}, the generalization of the result,  as suggested by model-dependent studies \cite{ollivier2},  allows for verifying Quantum Darwinism in realistic, imperfect ($\delta \neq 0$) scenarios. 

We observe that Eq.~(\ref{eq7}) holds  when ${\cal S}$ and {\it all} fragments of a certain size $k\geq k_{\delta}$ share a certain amount of classical correlations. The criterion can be relaxed by replacing $J\left({\cal S}:\check{\cal F}_{k}\right)$ with its average value over all ${\cal F}_k$. Under this less strong condition,  bounds like Eq.~(\ref{eq10})  exist for the average $I\left({\cal F}_{k,\hat{\mathbf{M}}_k}:{\cal F}_{l,\hat{\mathbf{M}}_l}\right)$. Also, adopting Eq.~(\ref{otherdarwin}) as Quantum Darwinism signature is justifiable {\it a posteriori}.  The quantity $J\left(\check{\cal S}:{\cal F}_k\right)$ displays the same scaling with $k$ of $J\left({\cal S}:\check{{\cal F}}_k\right)$ (and  $I({\cal S}:{\cal F}_k)$) in the widely applicable ``c-maybe'' model  in Fig.~\ref{fig2} \cite{akram}.
 
Finally, we show how to certify the emergence of classical objectivity when the universe is in a certain state $\ket{\psi}_{{\cal SE}}$. Verifying Eq.~(\ref{eq7}) 
is computationally hard, requiring an optimization over all possible measurements on ${\cal F}_k$  \cite{discorev,discome,corr1,corr2,thao}. The problem is bypassed  by linking  Quantum Darwinism to the scaling of an analytical function. Consider the conditional mutual information $I({\cal S}:{\cal F}_l\,|{\cal F}_k):=I({\cal S}:{\cal F}_{k+l})-I({\cal S}:{\cal F}_{k})$, which is  the supplemental information one acquires about ${\cal S}$ by enlarging the monitored fragment  \cite{cover,squash}. If and only if  such information is vanishing, then independent observers access maximal classical information about ${\cal S}$ \cite{epaps}:

 {\it {\bf Result 4:} For any state  $\ket{\psi}_{{\cal SE}}$, given $k_\delta \leq N/2,$}
\begin{align}
 \label{eq9}
  J\left({\cal S}:\check{\cal F}_k\right)&=\,H({\cal S}),\, \forall\,{\cal F}_k,\,  k\geq \, k_\delta\Rightarrow\nonumber\\
  I({\cal S}:{\cal F}_l|{\cal F}_k)&=0,\, \forall\,{\cal F}_k,{\cal F}_l,\, k \geq k_\delta, k+l\leq N-k_\delta \Rightarrow\nonumber\\
     J\left({\cal S}:\check{\cal F}_k\right)&=\,H({\cal S}),\, \forall\,{\cal F}_k,\,  k\geq 2\,k_\delta.
\end{align} 
Therefore, the Quantum Darwinism condition (\ref{eq7}) can be verified, in the strongest form ($\delta =0$), without explicit calculation of classical and quantum correlations. A more general one-way implication  reads
\begin{align}
\label{eq13}
J\left({\cal S}:\check{{\cal F}}_k\right)&\geq (1-\delta)\,H({\cal S}),\,\forall\, {\cal F}_k,\,\forall\,k\geq k_\delta\Rightarrow\\ 
I({\cal S}:{\cal F}_l|{\cal F}_k)&\leq 2\, \delta\,H({\cal S}),\forall\, {\cal F}_k,{\cal F}_l,\,k \geq k_\delta, k+l\leq N-k_\delta.\nonumber
\end{align} 
 Redundancy of classical information allows  the mutual information to increase rapidly only for $k>N-k_\delta$.  Quantum correlations, and therefore quantum information about ${\cal S}$, significantly build-up only in large fragments.   
 
\paragraph{Conclusion.} We have established universal, quantitative bounds on quantum correlations in multipartite systems.  
Independent observers   can  simultaneously access   classical information about a quantum system that redundantly spreads 
 into the environment, but quantum information is  out of reach. Hence, 
 {\it classical reality is not only consistent with  quantum laws, but an emergent byproduct of decoherence and Quantum Darwinism}.  We conjecture that  stronger bounds might exist when the environment state is mixed \cite{ZQZ1,ZQZ2},  and for multipartite correlations  \cite{cover,multi}. Also, the analytical witness of Quantum Darwinism may enable its experimental verification in large dimensional systems. 

\begin{acknowledgments}
{\it Acknowledgments.} 
This research was supported by grants FQXiRFP-1808 and FQXiRFP-2020-224322 from the Foundational Questions Institute and Fetzer Franklin Fund, a donor advised fund of Silicon Valley Community Foundation (SD and WHZ, respectively), as well as by the Department of Energy under the LDRD program in Los Alamos. A.T., B.Y. and W.H.Z. also acknowledge support from U.S. Department of Energy, Office of Science, Basic Energy Sciences, Materials Sciences and Engineering Division, Condensed Matter Theory Program, and the Center for Non- linear Studies. D. G. acknowledges financial support from the Italian Ministry of Research and Education (MIUR), grant number 54$\_$AI20GD01, and by a starting package of Politecnico di Torino, grant number 54$\_$RSG20GD01.
\end{acknowledgments}

\clearpage
\onecolumngrid
\renewcommand{\bibnumfmt}[1]{[A#1]}
 
\renewcommand{\citenumfont}[1]{{A#1}}

\setcounter{page}{1}
\setcounter{equation}{0}
 
\appendix*
\section{{\large Redundantly amplified information suppresses  quantum correlations in many-body systems}}  

\section{SUPPLEMENTARY MATERIAL}

\subsection*{Proofs of the technical results in the main text}\label{proofs}

 
{\it Justification of Eq.~(\ref{delta}) --} We prove that the parameter $\delta:=\frac{
J\left({\cal S}:\check{{\cal E}}\right)-\min\left\{J\left({\cal S}:\check{{\cal F}}_k\right),J\left({\cal S}:\check{{\cal E}}_{/k}\right)\right\}}{H({\cal S})}$ is a good measure of (lack of) classical objectivity. That is, it consistently evaluates  how different it is the information about  ${\cal S}$ that is accessible by measuring on ${\cal F}_k$ and ${\cal E}_{/k}$. From now on,
we assume with no loss of generality that $J\left({\cal S}:\check{{\cal E}}_{/k}\right)\geq J\left({\cal S}:\check{{\cal F}}_{k}\right)$. \\
First, if $\delta=0$, then $J\left({\cal S}:\check{{\cal E}}\right)=J\left({\cal S}:\check{{\cal E}}_{/k}\right)=J\left({\cal S}:\check{{\cal F}}_{k}\right)$. Also, the reverse implication is true. Therefore, the parameter vanishes if and only if there is perfect agreement between observers monitoring the two environment fragments.\\
Second, since $J\left({\cal S}:\check{{\cal E}}\right)= H({\cal S})$ for pure states of the universe, one has $\delta = 1- \frac{J\left({\cal S}:\check{{\cal F}}_k\right)}{H({\cal S})}$. Hence, if $\delta=1$, then $J\left({\cal S}:\check{{\cal F}}_k\right)=0$, as there is complete lack of consensus between the observers.  The reverse statement is also true. \\
Finally, we note that $\delta\rightarrow0$ in the degenerate case $H({\cal S})\rightarrow0$, as no information would be broadcast by ${\cal S}$ to the environment. \\

{\it Proof of Eq.~(\ref{main}), {\bf Result 2} --} We extend the upper bound to quantum discord to the multipartite case. We remind the definition of the parameters $
\delta_i:=\ \frac{
J\left({\cal S}:\check{{\cal E}}\right)-\min\left\{J\left({\cal S}:\check{\varepsilon}_i\right),J\left({\cal S}:\check{{\cal E}}_{/i}\right)\right\}}{H({\cal S})},$
 and their average  $\delta:=\sum_{i=1}^{N} \delta_i/N$. Then, quantum discord between ${\cal S}$ and an environment subsystem $\varepsilon_i$ is upper bounded as follows:
\begin{align*}
 I({\cal S}:\varepsilon_i)+ I({\cal S}:{\cal E}_{/i})&= 2\, H({\cal S})\Rightarrow\\
 D({\cal S}:\check{\varepsilon}_i)&= 2\, H({\cal S})- J({\cal S}:\check{\varepsilon}_i)-I({\cal S}:{\cal E}_{/i})\\
  D({\cal S}:\check{\varepsilon}_i)&\leq 2\, H({\cal S})+\delta_i\, H({\cal S})-H({\cal S})-I({\cal S}:{\cal E}_{/i})\\
  &\leq  H({\cal S})+\delta_i\, H({\cal S})-I({\cal S}:{\cal E}_{/i}).
 \end{align*}
Now, we note that for any pure state of the universe ${\cal SE}$,
 \begin{align*}
  I({\cal S}:\varepsilon_{i})+ I({\cal S}:{\cal E}_{/i})&= 2\, H({\cal S}),\,\forall\,i\Rightarrow\\
\nonumber
 I({\cal S}:\varepsilon_{i})+ I({\cal S}:{\cal F}_{k})&\leq 2\,H({\cal S}),\,\forall\, i,\,\forall\, {\cal F}_k\subseteq {\cal E}_{/i}\Rightarrow\nonumber\\
\bar I({\cal S}:{\cal E}_{/i}):=\frac1N  \sum_{i=1}^N  I({\cal S}:{\cal E}_{/i})&\geq H({\cal S}).
\end{align*}  
Therefore, the bound to the average quantum discord is  proven:
\begin{align*}
  D({\cal S}:\check{\varepsilon}_i)&\leq H({\cal S})+\delta_i\, H({\cal S})-I({\cal S}:{\cal E}_{/i})\Rightarrow\\
  \frac1N\sum_{i=1}^ND({\cal S}:\check{\varepsilon}_i)&\leq   \frac1N\sum_{i=1}^N\left\{ H({\cal S})+\delta_i\, H({\cal S})-I({\cal S}:{\cal E}_{/i})\right\}\\
  \bar D({\cal S}:\check{\varepsilon}_i)&\leq H({\cal S})+ \delta\, H({\cal S}) - \bar{I}({\cal S}:{\cal E}_{/i})\Rightarrow\\
    \bar D({\cal S}:\check{\varepsilon}_i)&\leq \delta\,H({\cal S}).
 \end{align*}


{\it Proof of Eq.~(\ref{main2}) --} For any pure state of the universe ${\cal SE}$, one has
 \begin{align*}
  I({\cal S}:\varepsilon_{i})+ I({\cal S}:{\cal E}_{/i})&= 2\, H({\cal S}),\,\forall\,i\Rightarrow\\
\nonumber
 I({\cal S}:\varepsilon_{i})+ I({\cal S}:{\cal F}_{k})&\leq 2\,H({\cal S}),\,\forall\, i,\,\forall\, {\cal F}_k\subseteq {\cal E}_{/i}\Rightarrow\nonumber\\
\bar I({\cal S}:\varepsilon_{i})&\leq H({\cal S})\\
\nonumber
 \bar D\left({\cal S}:\check\varepsilon_{i}\right)+\bar J\left({\cal S}:\check\varepsilon_{i}\right)&\leq H({\cal S}).
\end{align*}  
Hence, $\bar J\left({\cal S}:\check\varepsilon_{i}\right)\geq (1-\delta)\,H({\cal S}) \Rightarrow\bar D\left({\cal S}:\check\varepsilon_{i}\right)\leq \delta\,H({\cal S})$.\\

{\it Proof of Eq.~(\ref{redundancy}) --} Let us assume that  (at least) $R_{\delta}$ environment subsystems $\varepsilon_i$ share a certain amount of classical correlations with ${\cal S}$:
\begin{align*}
J\left({\cal S}:\check{\varepsilon}_i\right)&\geq(1-\delta)\,H({\cal S}),\,\# i=R_\delta\Rightarrow\\
\bar J\left({\cal S}:\check{\varepsilon}_i\right)&\geq \frac{R_\delta}{N}\,(1-\delta)\,H({\cal S})\Rightarrow\\
\bar J\left({\cal S}:\check{{\cal F}}_k\right)&\geq \frac{R_\delta}{N}\,(1-\delta)\,H({\cal S}),\,\forall\,k.
\end{align*}
  Since $\bar I({\cal S}:{\cal F}_k)\leq H({\cal S}),\,k\leq N/2$, the statement is proven:  $\bar{D}\left({\cal S}:\check{{\cal F}}_k\right)\leq \left(1-R_\delta\,(1-\delta)/N\right)\,H({\cal S}),\,k\leq N/2$.\\
  
  {\it Proof of Eq.~(\ref{discord}) -- } One has
 \begin{align*}
 J\left({\cal S}:\check{{\cal F}}_k\right)\geq&\, (1-\delta)\,H({\cal S}),\,\forall\,{\cal F}_k,\,k\geq k_\delta\Rightarrow\\
  I\left({\cal S}:{\cal F}_k\right)\geq&\, (1-\delta)\,H({\cal S}),\,\,\forall\,{\cal F}_k,\,k\geq k_\delta.
 \end{align*}
 Since $I\left({\cal S}:{\cal F}_k\right)+I\left({\cal S}:{\cal E}_{/k}\right)=2\,H({\cal S})$, one has $I\left({\cal S}:{\cal F}_{k}\right)\leq (1+\delta)\,H({\cal S}),\,\forall\,{\cal F}_{k},\,k\leq N-k_\delta$. Then, the statement is proven:
  \begin{align*}
  D\left({\cal S}:\check{{\cal F}}_k\right)\leq 2\,\delta\,H({\cal S}),\,\forall\, {\cal F}_k,\,k \in [k_\delta,N-k_\delta].
   \end{align*}

{\it Proof of Eq.~(\ref{eq10}), {\bf Result 3} --} Let us start by proving  the lower bound to $I\left({\cal F}_{k,\hat{\mathbf{M}}_k}:{\cal F}_{l,\hat{\mathbf{M}}_l}\right)$. One has
\begin{align*}
&J\left({\cal S}:\check{\cal F}_k\right)\equiv I\left({\cal S}:{\cal F}_{k,\hat{\mathbf{M}}_k}\right)\geq\,(1-\delta)\, H({\cal S}),\,\forall\,{\cal F}_k,\,k\geq k_\delta\Rightarrow\\
&I\left({\cal S}:{\cal F}_{l,\hat{\mathbf{M}}_l}\,\big|{\cal F}_{k,\hat{\mathbf{M}}_k}\right)=\, I\left({\cal S}:{\cal F}_{k,\hat{\mathbf{M}}_k}{\cal F}_{l,\hat{\mathbf{M}}_l}\right)-I\left({\cal S}:{\cal F}_{k,\hat{\mathbf{M}}_k}\right)\leq \delta\,H({\cal S}),\,\forall\,{\cal F}_k,{\cal F}_l,\,k\geq k_\delta\\
 &H({\cal S})+H\left({\cal F}_{k,\hat{\mathbf{M}}_k}{\cal F}_{l,\hat{\mathbf{M}}_l}\right)-H\left({\cal SF}_{k,\hat{\mathbf{M}}_k}{\cal F}_{l,\hat{\mathbf{M}}_l}\right)-H({\cal S})-H\left({\cal F}_{k,\hat{\mathbf{M}}_k}\right)+H\left({\cal S}{\cal F}_{k,\hat{\mathbf{M}}_k}\right)\leq\delta\,H({\cal S}),\,\forall\,{\cal F}_k,{\cal F}_l,\,k\geq k_\delta.
\end{align*}
Applying the strong subadditivity of the von Neumann entropy,
\begin{align*}
 -H\left({\cal S}{\cal F}_{k,\hat{\mathbf{M}}_k}{\cal F}_{l,\hat{\mathbf{M}}_l}\right)-H({\cal S})+H\left({\cal S}{\cal F}_{k,\hat{\mathbf{M}}_k}\right)\geq\,  -H\left({\cal S}{\cal F}_{l,\hat{\mathbf{M}}_l}\right).
\end{align*}
Hence, 
\begin{align*}
& H\left({\cal F}_{k,\hat{\mathbf{M}}_k}\right)+H\left({\cal S}{\cal F}_{l,\hat{\mathbf{M}}_l}\right)-H\left({\cal F}_{k,\hat{\mathbf{M}}_k}{\cal F}_{l,\hat{\mathbf{M}}_l}\right)\geq\, (1-\delta)\,H({\cal S}),\,\forall\,{\cal F}_k,{\cal F}_l,\,k\geq k_\delta\Rightarrow\\
&I\left({\cal F}_{k,\hat{\mathbf{M}}_k}:{\cal F}_{l,\hat{\mathbf{M}}_l}\right)\geq\, I\left({\cal S}:{\cal F}_{l,\hat{\mathbf{M}}_l}\right)-\,\delta\,H({\cal S}),\,\forall\,{\cal F}_k,{\cal F}_l,\,k\geq k_\delta \Rightarrow \\
 &I\left({\cal F}_{k,\hat{\mathbf{M}}_k}:{\cal F}_{l,\hat{\mathbf{M}}_l}\right)\geq\, \left(1-2\,\delta\right)H({\cal S}),\,\forall\,{\cal F}_k,{\cal F}_l,\,k,l\geq k_\delta.
\end{align*}
The lower bound to $I\left({\cal F}_{k,\hat{\mathbf{M}}_k}:{\cal F}_{l,\hat{\mathbf{M}}_l}\right)$ is then proven  for any pure state $\ket{\psi}_{{\cal SE}}$, whenever the Quantum Darwinism condition (Eq.~(12) of the main text) is verified. \\
Now, we focus on states of the form ${\bf U}_{{\cal SE}}\ket{\psi}_{{\cal S}}\ket{\phi}_{{\cal F}_k}\ket{\varphi}_{{\cal F}_l}\ket{\chi}_{{\cal E}_{/k+l}}$, with ${\bf U}_{{\cal SE}}\equiv{\bf U}_{{\cal SE}_{/k+l}}{\bf V}_{{\cal SF}_l}{\bf W}_{{\cal SF}_k}$. We quantify  by $\Delta_{{\bf U}} X$ the difference between the values of a quantity $X$, as computed after and before applying a unitary map ${\bf U}$. 
Being $H({\cal S})$  the entropy of the final state of ${\cal S}$, given the hypothesis $\Delta_{{\bf U}_{{\cal SE}}}I\left({\cal S}:{\cal F}_{k,\hat{\mathbf{M}}_k}\right)\geq (1-\delta)\,H({\cal S}),\,\forall\,{\cal F}_k,\, k\geq k_\delta$,
we find
\begin{align*}
\Delta_{{\bf U}_{{\cal SE}}}I\left({\cal F}_{k,\hat{\mathbf{M}}_k}:{\cal F}_{l,\hat{\mathbf{M}}_l}\right)&\leq \Delta_{{\bf U}_{{\cal SE}}}I\left({\cal SF}_{k,\hat{\mathbf{M}}_k}:{\cal F}_{l,\hat{\mathbf{M}}_l}\right),\,\forall\,{\cal F}_k,{\cal F}_l,\forall\,k,l\\
&\leq \Delta_{{\bf U}_{{\cal SE}}}I\left({\cal S}:{\cal F}_{l,\hat{\mathbf{M}}_l}\right)+ \Delta_{{\bf U}_{{\cal SE}}}I\left({\cal F}_{k,\hat{\mathbf{M}}_k}:{\cal F}_{l,\hat{\mathbf{M}}_l}\big|{\cal S}\right),\,\forall\,{\cal F}_k,{\cal F}_l,\forall\,k,l\\
&\leq H({\cal S})+\Delta_{{\bf U}_{{\cal SE}}}I\left({\cal F}_{k,\hat{\mathbf{M}}_k}:{\cal SF}_{l,\hat{\mathbf{M}}_l}\right)-\Delta_{{\bf U}_{{\cal SE}}}I\left({\cal S}:{\cal F}_{k,\hat{\mathbf{M}}_k}\right),\,\forall\,{\cal F}_k,{\cal F}_l,\forall\,k,l\\
&\leq \delta\,H({\cal S})+\Delta_{{\bf U}_{{\cal SE}}}I\left({\cal F}_{k,\hat{\mathbf{M}}_k}:{\cal SF}_{l,\hat{\mathbf{M}}_l}\right),\,\forall\,{\cal F}_k,{\cal F}_l,\,k\geq k_\delta, \forall\,l\\
&\leq \delta\,H({\cal S})+\Delta_{{\bf W}_{{\cal SF}_k}}I\left({\cal F}_{k,\hat{\mathbf{M}}_k}:{\cal SF}_{l,\hat{\mathbf{M}}_l}\right),\,\forall\,{\cal F}_k,{\cal F}_l,\,k\geq k_\delta, \forall\,l\\
&\leq\delta\,H({\cal S})+\Delta_{{\bf W}_{{\cal SF}_k}}I\left({\cal S}:{\cal F}_{k,\hat{\mathbf{M}}_k}\right),\,\forall\,{\cal F}_k,{\cal F}_l,\,k\geq k_\delta, \forall\,l\\
&\leq \begin{cases}
 \left(1+\,\delta\right)\,H({\cal S}),\,\,
    \text{if}\,\Delta_{{\bf U}_{{\cal SE}_{/k+l}}{\bf V}_{{\cal SF}_l}}H({\cal S})\geq 0,\\
    \log_2 d_{{\cal S}}+\delta\,H({\cal S}), \, \text{otherwise},\,\forall\,{\cal F}_k,{\cal F}_l, \,k\geq k_\delta, \forall\,l
\end{cases}
\Rightarrow\\
&\leq \begin{cases}
 \left(1+\,\delta\right)\,H({\cal S}),\,\,
    \text{if}\,\Delta_{{\bf U}_{{\cal SE}_{/k+l}}{\bf V}_{{\cal SF}_l}}H({\cal S})\geq 0,\\
    \log_2 d_{{\cal S}}+\delta\,H({\cal S}), \, \text{otherwise},\,\forall\,{\cal F}_k,{\cal F}_l, \,k,l\geq k_\delta.
\end{cases}
\end{align*}
 Thus,   the upper bound   is proven as well.\\

{\it Proof of Eq.~(\ref{eq9}), {\bf Result 4} --} Let us start by proving the first  ``$\Rightarrow$'' relation. By applying the Koashi-Winter relation, one has
\begin{align*}
J\left({\cal S}:\check{\cal F}_k\right)&=H({\cal S}),\forall\,{\cal F}_k,\,k\geq k_\delta\Rightarrow\\
 E_f({\cal S}:{\cal E}_{/k})&=D\left({\cal S}:\check{\cal F}_k\right)+H({\cal S}\,|{\cal F}_k)=0,\,\forall\,{\cal F}_k,\,k\geq k_\delta\Rightarrow\\ D\left({\cal S}:\check{\cal F}_k\right)&=0, \forall\,{\cal F}_k,\,k\in[k_\delta,  N-k_\delta].
\end{align*}
The second implication holds because  there  is no entanglement between ${\cal S}$ and any fragment ${\cal E}_{/k}$ of size  $N-k\leq N-k_\delta$, so the conditional entropy $H({\cal S}\,|{\cal F}_k)$ is never negative for $k\leq N-k_\delta$. Note that we are assuming that $k_\delta\leq N/2$. Then,  both the conditional entropy and quantum discord must be exactly zero for $k\in[k_\delta, N-k_\delta]$.
In conclusion, $I({\cal S}:{\cal F}_k)= H({\cal S}),\,\forall\, k\in[k_\delta, N- k_\delta]$, and  therefore
\begin{align*}
I({\cal S}:{\cal F}_{k+l})&=I({\cal S}:{\cal F}_k)=  H({\cal S}), \, \forall\,{\cal F}_k,{\cal F}_l,\,k \geq k_\delta, k+l\leq N-k_\delta\Rightarrow\\
  I({\cal S}:{\cal F}_l\,|{\cal F}_k)&=0,\,\, \forall\,{\cal F}_k,{\cal F}_l,\,k \geq k_\delta, k+l\leq N-k_\delta.
\end{align*}
We now demonstrate the second ``$\Rightarrow$'' relation.
 The hypothesis is  $I({\cal S}:{\cal F}_l|{\cal F}_k)=0,\, \forall\,{\cal F}_k,{\cal F}_l,\,k \geq k_\delta, k+l\leq N-k_\delta$. The condition signifies that every bipartition ${\cal SF}_l,\,l\leq N-2\,k_\delta$  is in a separable state, as the conditional mutual information $I({\cal S}:{\cal F}_l|{\cal F}_k)$ is an upper bound to the squashed entanglement in ${\cal SF}_l$ (see Ref.[68] of the main text). By employing again the Koashi-Winter relation, we have
\begin{align*}
J\left({\cal S}:\check{\cal F}_k\right)&= H({\cal S})-E_f({\cal S}:{\cal E}_{/k}),\,\forall\,{\cal F}_k\Rightarrow\\
J\left({\cal S}:\check{\cal F}_k\right)&= H({\cal S}),\,\forall\,{\cal F}_k,\,k \geq 2\,k_\delta.
\end{align*}

{\it Proof of Eq.~(\ref{eq13}) --} 
 While proving Eq.~(\ref{discord}), we have demonstrated that Quantum Darwinism dictates that   $I\left({\cal S}:{\cal F}_k\right)\leq (1+\delta)\,H({\cal S}),\,k\leq N-k_\delta$. Thus, 
  \begin{align*}
  I\left({\cal S}:{\cal F}_l|{\cal F}_k\right)\leq 2\,\delta\,H({\cal S}),\,\forall\, {\cal F}_k,{\cal F}_l,\,k \geq k_\delta, k+l\leq N-k_\delta.
   \end{align*}

\subsection*{Full details on the case study in Fig.~\ref{fig2} of the main text} 
  We consider a qubit ${\cal S}$ and an $N$-qubit  environment ${\cal E}$ in the initial state $\ket{+}_{{\cal S}} \ket{0}^{\otimes N}_{{\cal E}}$, with $\ket{+}\equiv \sqrt{1/2}(\ket{0}+\ket{1})$. The universe undergoes the evolution ${\bf U}_{{\cal SE}}(a):=\Pi_{i=1}^N{\bf U}(a)_{{\cal S}\varepsilon_{i}}$. Every unitary map ${\bf U}_{{\cal S}\varepsilon_{i}}(a)$ implements a controlled operation 
$\mathbb{I}_2\oplus  \left( \begin{smallmatrix} a&\sqrt{1-a^2}\\ \sqrt{1-a^2}&-a \end{smallmatrix} \right)$ in the bipartition ${\cal S}\varepsilon_{i}$, such that ${\cal S}$ is the control qubit. These c-maybe gates  are generated by a sequence of commuting Hamiltonians $H_{{\cal SE}}=\sum_{i=1}^N H_{{\cal S}\varepsilon_i}$.  

The system initial state and the global unitary map are manifestly invariant under swapping of the environment components $\varepsilon_i$. Therefore, the average values of quantum and classical correlations correspond to the values we compute on an arbitrary ${\cal S}\varepsilon_i$ bipartition.
By partial trace of the final state ${\bf U}_{{\cal SE}}(a)\ket{+}_{{\cal S}} \ket{0}^{\otimes N}_{{\cal E}}$, one finds that the marginal state of ${\cal S}\varepsilon_i$ is a rank-two density matrix, for any $N$. In such a case, it is possible to find analytical expressions for quantum discord and classical correlations (see Refs.[58,59] of the main text),   which read:
\begin{align*}
h(x):&=-\frac{1-x}{2}\,\log_2\frac{1-x}{2}-\frac{1+x}{2}\,\log_2\frac{1+x}{2},\\
\bar D\left({\cal S}:\check{\varepsilon}_i\right)&=h(a)-h\left(a^{N-1}\right)+h\left(\sqrt{a^{2N}-a^2+1}\right), \\
\bar J\left({\cal S}:\check{\varepsilon}_i\right)&= h\left(a^N\right)-h\left(\sqrt{a^{2 N}-a^2+1}\right).  
\end{align*}
Also,  as they are instrumental to compute the upper bound in Eq.~(\ref{main}) and they are plotted in Fig.~\ref{fig2}, we write down the following expressions:
\begin{align*}
H({\cal S})&=h\left(a^N\right),\\
H(\varepsilon_i)&=h\left(a\right),\forall\,i,\\
\delta=\delta_i &=\frac{h\left(\sqrt{a^{2 N}-a^2+1}\right)}{h\left(a^N\right)},\,\forall\,i. 
\end{align*}

\begin{thebibliography}{99}
\bibitem{notent}The  entanglement of formation of $\rho_{{\cal SF}_k}=\sum_\alpha p_\alpha\rho_{\alpha,{\cal SF}_k}$, with $\rho_{\alpha}=\ket{\alpha}\bra{\alpha}$,  is  $E_f({\cal S}:{\cal F}_k):=\min\limits_{\left\{p_\alpha,\rho_\alpha\right\}} -\sum_\alpha p_\alpha \text{tr}\left\{ \text{tr}_{{\cal S}}\left\{\rho_{\alpha,{\cal SF}_k}\right\}\log_2\text{tr}_{{\cal S}}\left\{\rho_{\alpha,{\cal SF}_k}\right\}\right\}$.
\bibitem{notealt}One has $\delta \leq 1-\frac{\min_i J\left({\cal S}:\check{\varepsilon}_i\right)}{H({\cal S})}$. The right-hand side could be an alternative objectivity quantifier.
\bibitem{note}The terms ``measurement'' and ``observable'' are used interchangeably, as experimental read-outs - the eigenvalue spectra of physical quantities - do not play any role in this discussion.
\bibitem{noteclass}The quantity  $J\left(\check{{\cal S}}:{\cal F}_k\right)$ is instead the maximal classical information about ${\cal F}_k$ that one accesses via measurements on ${\cal S}$ \cite{vedral}.
\end{thebibliography}
\end{document}